# Cross-over between channeling and pinning at twin boundaries in YBa$_2$Cu$_3$O$_7$ thin films

A. Palau[1], J.H. Durrell[1], J. L. MacManus-Driscoll[1], S. Harrington[1], T. Puig[2], F. Sandiumenge[2], X. Obradors[2] and M. G. Blamire[1]

[1]Department of Materials Science, University of Cambridge, Pembroke Street, Cambridge CB2 3QZ, U.K.

[2]Institut de Ciència de Materials de Barcelona, CSIC, Campus de la UAB, 08193 Bellaterra, Spain

The critical current ($J_c$) of highly twinned YBa$_2$Cu$_3$O$_7$ films has been measured as a function of temperature, magnetic field and angle. For much of the parameter space we observe a strong suppression of $J_c$ for fields in the twin boundary (TB) directions; this is quantitatively modeled as flux-cutting-mediated vortex channeling. For certain temperatures and fields a cross-over occurs to a regime in which channeling is blocked and the TBs act as planar pinning centers so that TB pinning enhances the overall $J_c$. In this regime, intrinsic pinning along the TBs is comparable to that between the twins.

A variety of different explanations have been proposed for the relatively high critical current densities ($J_c$s) found in YBa$_2$Cu$_3$O$_7$ (YBCO) thin films. These include dislocations [1], low angle grain boundaries [2], intrinsic pinning by the *a-b* planes [3,4], twin boundaries [5], stacking faults [4] and random defects associated to oxygen or cation vacancies [6]. Although it is often difficult to differentiate unambiguously the contributions to pinning from these different sources, it is possible to exploit their different temperature, field and anisotropy dependencies; for example, Civale *et al.* have shown that out-of-plane angular measurements can distinguish between different pinning defects (random vs. correlated) [7]. In the work described here we have exploited the fact that in epitaxial films twin boundaries always occur in particular directions in order to study their contributions to the $J_c$.

The role of twin boundaries in YBCO has been extensively studied but a full understanding of their influence on vortex pinning is still lacking. Theoretical calculations have variously predicted both an increase [8] and a decrease [9,10] of the superconducting order parameter at the twin boundaries. Although both possibilities are expected to create effective pinning sites for vortices, a decreased order parameter in the twin planes would be expected to give rise channeling of vortices along the twin boundaries, by analogy with the behavior of grain boundaries [11]. Gurevich and Cooley [12] discuss the general behavior of vortices interacting with planar defects and demonstrate theoretically that channeling should generally be favored because of the low longitudinal (i.e. within the plane) pinning force acting on vortices lying within such defects.

Depending on the type of measurement performed, previous experiments have shown the apparently contradictory results that twin planes can act as strong pinning centers [5,13-15] or as vortex channels [16-19]. In this Letter we present clear evidence from angular transport critical current measurements that twin boundaries can, in the same sample, act both as flux channels and as pinning centers, with a temperature- and magnetic field- dependent cross-over between these two regimes. In particular we show that under certain circumstances the longitudinal pinning force within a twin boundary is at least equal to that between the plains, thus blocking vortex channeling.

High $J_c$ YBCO films 150nm thick, were grown on LaAlO$_3$ single crystal substrates by chemical solution deposition (CSD) using an established trifluoroacetate (TFA) precursor [20]. These films were patterned to give 500μm long, 20μm width current tracks using optical lithography and argon ion milling. Tracks were patterned in different directions in the *a-b* plane in order to vary the angle of the maximum and minimum values of the Lorentz force with respect to the YBCO crystallographic orientations. In twinned films, both [100] and [010] directions in YBCO are parallel to each of the [100] and [010] directions in the substrate plane. Thus we consider the orientation of the track with respect the [100] and [010] directions, denoted by $\alpha$ (see inset of Fig. 1). Tracks were patterned with $\alpha=0^o$ in samples A and B and with $\alpha=25^o$ and $\alpha=80^o$ in samples C and D, respectively.

In order to determine the effect of directional defects at different temperatures and magnetic fields, critical current density, $J_c$, data were obtained by rotating the magnetic field in-plane ($\phi$) and out-of-plane ($\theta$) of the film surface, using a two-axis goniometer mounted in an 8 T magnet [21]. Critical currents were determined using a voltage criterion of 0.5 µV. Figure 1 shows the $J_c(\phi)$ curves obtained at $\theta=60°$, $T = 50$ K and $\mu_0 H = 5$ T for tracks A, C and D. The broad peaks in $J_c$ observed at $\phi=0°$ and $180°$ are those expected in the minimum Lorentz force configuration. Superimposed on this in all the curves in Fig.1 is a sharp suppression of $J_c$ when the magnetic field is oriented along the [110] and [$\bar{1}$10] YBCO directions. The rapid increase of $J_c(\phi)$ on either side of the minima resembles that previously observed [11,22] where vortex channeling occurs along grain boundaries. For sample A, with $\alpha=0°$, the minima occur at $\phi= -45°, 45°, 135°$, etc, while for the samples with $\alpha\neq 0°$, the minima are correspondingly shifted according to the value of $\phi-\alpha$. Note that the suppression of $J_c$ in the channeling direction is larger when the twin orientation lies close to the angle of the maximum Lorentz force ($\phi=90°$) and is less pronounced when the twin orientation lies near the angle of the minimum Lorentz force ($\phi=0°$ and $180°$).

Figure 2 shows the effect of changing the out-of-plane angle of the applied magnetic field, $\theta$, on the $J_c(\phi)$ behavior for sample C at $\mu_0 H=5$T for two different temperatures, $T = 50$K and $T = 77$K. At $T = 50$K (Fig. 2(a)) we see vortex channeling minima along the [110] and [$\bar{1}$10] orientations for all the curves measured, although the depths of the minima are reduced at high values of $\theta$. The inset to Fig. 2(a) shows several $J_c(\phi)$ curves measured at different values of $\theta$ ($\theta>60°$) for sample A at 50K and $\mu_0 H=7$T. The channeling minimum is visibly reduced as the value of $\theta$ is increased and, indeed, no channeling at all can be detected for $\theta=90°$ (magnetic field parallel to the film plane). In contrast, at 77K (Fig. 2(b)) the $J_c(\phi)$ curves for $\theta=20°$ and $\theta=30°$ show broad peaks instead of the channeling minima at the same $\phi$ orientations. In fact, such peaks can also be observed for the other values of $\theta<90°$, although for $\theta>40°$ they are superimposed on the channeling minima and merge into the minimum Lorentz force peak as $\theta\rightarrow 90°$. The appearance of broad $J_c$ peaks indicates that at high temperatures, i.e. lower overall pinning energies, the planes that induced vortex channeling along the [110] and [$\bar{1}$10] orientations also generate directional vortex pinning. The maximum $J_c$ peaks are observed at low $\theta$ values when a large component of the Lorentz force is normal to the twin plane and a correspondingly small component lies along the channel.

Figure 3 shows several $J_c(\phi)$ curves measured at different temperatures for sample C at $\mu_0 H=5$T and $\theta=60°$. Values of $J_c$ have been normalized to the value at $\phi=90°$ in order to distinguish the evolution of the channeling minimum and the $J_c$ peak with temperature. At low temperatures (T=20K and 50K) we only observe the channeling minima whereas at higher temperatures (T=77K) a tiny broad peak superimposed on the minima can be distinguished, especially near the maximum Lorentz force ($\phi=70°$). At T=80K the $J_c$ peaks superimposed on the channeling minima are clearly visible at $\phi=-20°$, $\phi=70°$ and $\phi=160°$. The inset to Figure 3 shows several $J_c(\phi)$ curves measured at 50K and $\theta=60°$ for sample A at different applied magnetic fields. We observe that the channeling effect appears for $H > 2$ T and the channeling minimum becomes more prominent with increasing magnetic field. The presence of $J_c$ peaks observed at lower temperatures has also been observed for $H > 2$ T.

Figure 4 shows a TEM planar view image of a typical YBCO film grown by TFA which reveal the existence of a high biaxial twin density with typical domain spacing of 50 nm, containing either [110] or [$\bar{1}$10] TBs. We have shown that both directional vortex pinning and channeling effects occur along these directions, depending on the temperature, magnetic field or out-of-plane angle, $\theta$, and these effects can be associated with the high density of TBs present in the YBCO thin films studied. To confirm the generality of this behavior we have studied YBCO thin films grown on $SrTiO_3$ substrates by pulsed laser deposition (PLD), and obtained similar results. TEM planar view images of YBCO thin films grown by PLD show that they also present a high density of twin boundaries [23]. Thus, the effects reported here can be treated as a general result which occurs for all the twinned YBCO films.

For vortex channeling to be observable three requirements must be satisfied, (1) vortices must be confined to the planar defects by a lower local value of the order parameter, (2) these defects must be preferentially aligned within the sample, and (3) the longitudinal pinning along the boundary must be lower than that within the twin domains. For (3) we make the important qualification that, in general, the longitudinal pinning force is actually the sum of the longitudinal elementary pinning force $f_\parallel$ introduced by Gurevich and Cooley [12] and the cutting force $f_{cut}$ required for vortices at an arbitrary angle to the defect to flow along it. It is the increase in this cutting force and hence in the total longitudinal pinning with increasing angle between the vortices and the defect which gives rise to the cross-over previously observed in grain boundaries where, for larger angles, flux flow occurs within the grains rather than by channeling down the grain boundaries [11].

By analogy with grain boundaries, we treat the flux channel within the TB as a region where flux cutting can occur [11] and so write the Lorentz force balance equation as, $\mathbf{J_c} \times \mathbf{B} = \mathbf{F}_\Box + \mathbf{F}_{cut}$, where $\mathbf{F}_\Box$ is the pinning force density of the vortex segments within the twin boundary and $\mathbf{F}_{cut}$ is a summation of the flux cutting force over all vortices intersecting the twin boundary. $\mathbf{F}_\Box$ can be written in terms of the pinning force per unit length within the twin boundary, $f_\Box$, and the vortex spacing, $a_0$, as $|\mathbf{F}_\Box|=(f_\Box/a_0^2)$. Similarly $\mathbf{F}_{cut}$ can be written in terms of the force required to cut and cross-join [24] a single vortex, $f_{cut}$, and the density of points where vortices cross into the twin boundary, $|\mathbf{F}_{cut}|=(f_{cut}/L\, a_0^2)$, where $L$ is the length of each vortex segment in the twin boundary, given by $L=d_{tb}/|\cos(\phi-\alpha\pm 45)|$ and $d_{tb}$ is the twin boundary width. In the experimental geometry employed $|\mathbf{J}_c \times \mathbf{B}| = J_c B \sqrt{\sin^2\phi + \cos^2\phi \cos^2\theta}$, and thus the critical current density can be written as

$$J_c = \frac{f_\Box + \left(f_{cut}\left|\cos(\phi-\alpha\pm 45°)\right|/d_{tb}\right)}{\Phi_0 \sqrt{\sin^2\phi + \cos^2\phi \cos^2\theta}} \quad (1)$$

Equation 1 has been used for fitting the experimental $J_c(\phi)$ data in Fig. 1 near the channeling minima. For that, we have taken a value of $d_{tb}$=5nm, typical for YBCO [25], and $f_\Box$ and $f_{cut}$ left as fitting parameters. Solid lines in Fig. 1 shows the fitting curves obtained with $f_\Box$=5.4.10$^{-6}$ N/m, 8.4.10$^{-6}$ N/m and 8.9.10$^{-6}$ N/m and $f_{cut}$=2.1.10$^{-14}$ N, 4.8.10$^{-14}$ N, 2.7.10$^{-14}$ N for samples A, C and D, respectively. The slightly lower value of $f_\Box$ value for sample A is consistent with its lower overall critical current density. The $f_{cut}$ values found are very similar and the small differences could be due to different $d_{tb}$. Using the optimum values of $f_\Box$ and $f_{cut}$, the theoretical equation accurately reproduces the shape of the channeling minima measured for the three samples studied; including the increased depth of the $J_c$ minima close to the maximum Lorentz force. The same equation has been used to fit the evolution of the $J_c$ channeling minima with the out-of-plane angle, $\theta$. The inset to Fig 2(b) shows the experimental values of the $J_c$ minima obtained at $T = 50$ K and $\mu_0 H = 5$ T, for samples A and C at different $\theta$ values. Dashed lines show the $J_c(\theta)$ dependence determined by using equation 1 at the different $\phi$ values considered; the results fit well to the experimental data for $\theta \leq 80°$. For $\theta=90°$, i.e. magnetic field parallel to the film plane, the effect of intrinsic pinning in the $a$-$b$ planes is to increase the pinning force significantly and so the experimental $J_c$ is much higher than that predicted by equation 1.

The complete absence of channeling minima for $\theta=90°$ (see Fig. 2(b) and the inset to Fig. 2(a)) indicates that intrinsic pinning by the $a$-$b$ planes blocks the vortex channeling through the channel. This is very different to the behavior observed in GBs [11] and is presumably a consequence of a near perfect structural order of the YBCO within the twin boundaries. We observe TB channeling for a wide range of conditions and so conditions (1) and (2) are clearly satisfied in our experiments. Condition (3) has been considered previously by Gurevich and Cooley [12], and they conclude that for general planar structures in which modified Abrikosov-Josephson (AJ) vortices lie within the defect, $f_\Box \ll f_A$ where $f_A$ is the bulk pinning force per unit length of the superconductor, implying that channeling should be generally observed. Since, for $\theta \Box 90°$, channeling is blocked, we can conclude that the order parameter is not homogeneously low within the twin boundary, but rather preserves a similar spatial variation associated with the layered crystal structure to that in defect-free material. This is consistent with the model of Belzig *et al.* [9], in which an ideal twin boundary is non-pair breaking but has a suppressed order parameter over a distance much less than the coherence length. Thus the energy of a vortex, averaged over the extent of the core, would be reduced at a twin boundary, but would still be modulated by the periodic potential associated with the layered crystal structure within the material surrounding the boundary. This further implies that TBs cannot be considered to be 'weak links' under any circumstances and that the vortices retain their Abrikosov nature throughout the material.

The other major experimental observation reported here is that there are more general conditions of field strength and orientation (i.e. $\theta \neq 90°$,) under which channeling is either blocked or superimposed on a $J_c$ enhancement. This is most apparent at high temperature and low $\theta$ (see Fig 2(b)). This is a consequence of the particular symmetry of TBs: although, in general, components of the Lorentz force act both along and normal to both planes defining the TB directions, except in the special case of $\theta = 0$ vortices will only interact significantly with the TB plane to which they are more nearly parallel.

Where the $f_\Box < f_A$, channeling will determine $J_c$ for the material as a whole since vortices away from the TB experiencing the same Lorentz force will be more strongly pinned. If $f_\Box > f_A$ then the channel is invisible once vortex flow within untwinned material will occur, as described above for the special case of $\theta=90°$. The situation is different for Lorentz force components normal to the twin boundary; here the two-dimensional nature of the defect means that there is no short-circuit pathway for vortex flow and so any pinning associated with twin boundaries which exceeds that of the intervening material will enhance the overall $J_c$.

The inset diagram to Fig. 4 shows schematically a vortex interacting with a set of parallel planar defects. We discuss this interaction using a simplified version of the model introduced by Blatter *et al.* [26]. Here, the fraction $p$ of each vortex lying within the planar defect and hence able to interact with it is $p=|\cos\gamma|-|\sin\gamma/\tan\gamma'|$ and the energy per unit length is given by

$$E = \sin\gamma\left\{\frac{F}{\sin\gamma'} + (F-Z)|\cot\gamma - \cot\gamma'|\right\} \quad (2)$$

where $F$ is the vortex energy per unit length (which can be held to include general elasticity terms for the lattice as a whole), $Z$ is the energy reduction per unit length for a vortex within the planar defect (physically $F > Z$). Minimizing $E$ with respect to $\gamma'$ gives $\cos\gamma' = 1-Z/F$. Thus, as $Z$ tends to $F$ (i.e. the strong pinning limit), $p$ tends to the geometrical limit of $\cos\gamma$. As $Z$ decreases, $p$ declines increasingly rapidly with $\gamma$ and is zero, contributing no extra pinning, for $\gamma > \cos^{-1}(1-Z/F)$.

Taking, for simplicity, the dependence of the planar pinning as $|\cos\gamma|$, the behavior given by equating the in-plane component of the Lorentz force to $p$ is shown in the inset to Fig. 4. The same angles as Fig. 2(b) have been used and, without being a quantitative fit, the results reproduce well the general shapes and trends observed in the experimental data. Since, the characteristic angular dependence of the channeling means that a depressed $J_c$ occurs only within a few degrees of the TB, where it occurs simultaneously with enhanced TB pinning, this is observed as a sharp dip within a broad pinning peak. It is therefore possible for twin boundaries to act simultaneously as vortex channels and centers for enhanced 2-D pinning. Significantly, there is a regime at high temperatures (for example in Fig 2(b)), when bulk pinning is relatively weak, in which both the perpendicular and longitudinal pinning due to TBs exceeds that of the bulk and hence determines the overall critical current of the material irrespective of field orientation.

In conclusion, we have investigated the $J_c(\phi)$ dependence in YBCO thin films at different temperatures and magnetic fields. Previous studies of the effect of twin boundaries on pinning in YBCO have attributed both suppression and enhancement of $J_c$ to twin boundaries. By reducing the symmetry of the experimental measurement, we have shown that twin planes may act both as pinning sites or flux channels depending on the temperature and magnetic field orientation. The $J_c(\phi)$ dependence in the channeling minima resembles the one observed in low-angle grain boundaries [11] and vicinal films [22] and can be described by a flux cutting model. The disappearance of channeling for $\theta = 90°$ implies that vortices within TBs, but not other planar defects, remain strongly influenced by intrinsic pinning and so retain their Abrikosov nature.

This work was supported by national funds (EPSRC-UK and MAT2005-02047-Spain) and European Union (HIPERCHEM project, NMP4-CT2005-516858). A. Palau wishes to thank the government of Catalonia for a Beatriu de Pinós contract.

FIG. 1.

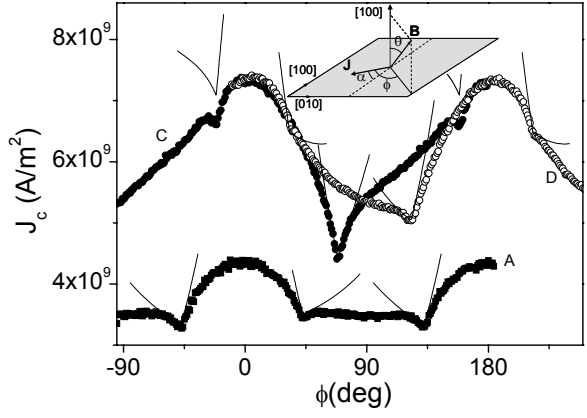

FIG. 2.

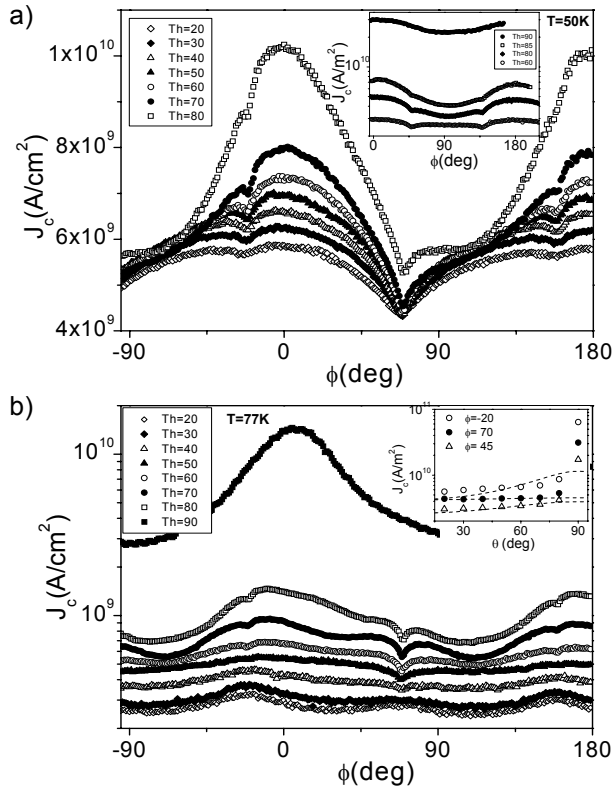

FIG. 3.

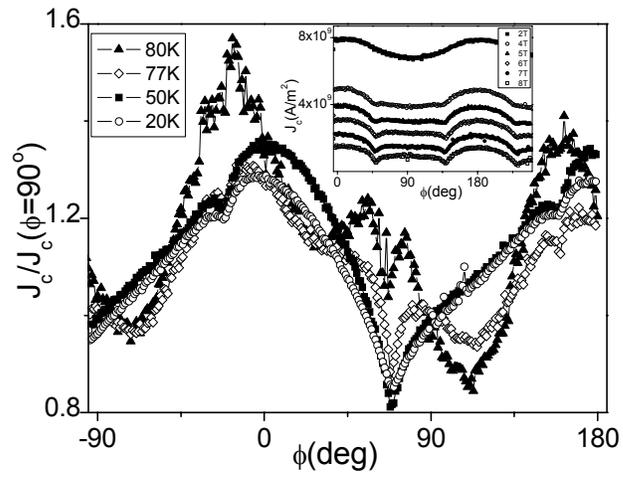

FIG. 4.

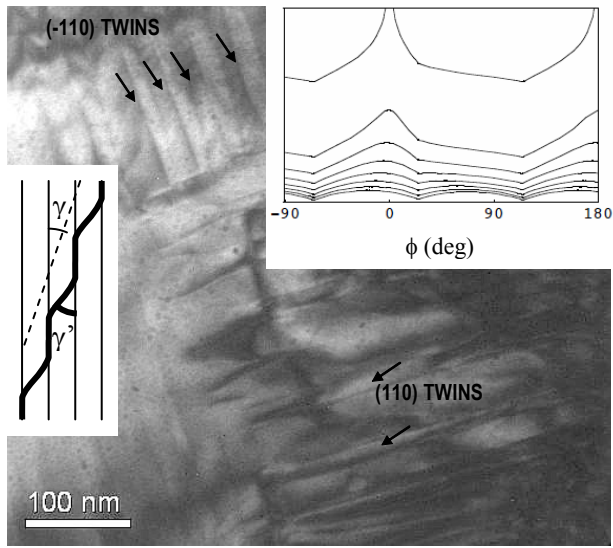

FIG. 1. Critical current dependence on the on the in-plane angle between the current and magnetic field, $J_c(\phi)$, measured at $\theta=60°$, $T = 50$ K and $\mu_0H = 5$ T for samples with different angles $\alpha$ between the current and (100) crystallographic direction A (0°), C (25°) and D (80°). Solid lines are theoretical fits to the experimental data using equation 1. The inset shows the experimental geometry considered.

FIG. 2. $J_c(\phi)$ dependence obtained for sample C at various $\theta$ angles for $\mu_0H = 5$ T at $T=50K$ (a) and 77K (b). The Inset to Fig. 2(a) shows several $J_c(\phi)$ curves measured at 50K and 7T for sample A, varying $\theta$ values with $\theta>60°$. The inset to Fig. 2(b) shows the $\theta$ dependence of the $J_c$ in the channeling minima at $T=50K$ and $\mu_0H = 5$ T, for sample A at $\phi=45°$ (△) and for sample C at $\phi=-20°$ (○) and $\phi=70°$ (●).

FIG. 3. $J_c(\phi)$ curves measured at different temperatures for sample C at 5T and $\theta=60°$, normalized to their value at $\phi=90$. The $J_c(\phi=90°)$ values obtained were $1.54.10^{10} A/m^2$, $5.44.10^9 A/m^2$, $5.25.10^8 A/m^2$, $1.98.10^8 A/m^2$ at 20K, 50K 77K and 80K respectively. Inset shows the $J_c(\phi)$ dependence obtained for sample A for various applied magnetic fields at 50K and $\theta=60°$.

FIG. 4. TEM planar view image of a typical YBCO film grown by TFA where a high density of twin boundaries along the [110] and [$\bar{1}$10] directions can be observed. The inset graph shows $J_c$s arising from TB pinning for the range of angles in Fig. 2(b) simulated using the model described in the text and illustrated by the inset diagram.

1. B. Dam *et al.*, Nature **399,** 439 (1999).

2. A. Diaz, L. Mechin, P. Berghuis, and J. E. Evetts, Physical Review Letters **80,** 3855 (1998).

3. B. Roas, L. Schultz, and G. Saemannischenko, Physical Review Letters **64,** 479 (1990).

4. L. Civale *et al.*, Physica C-Superconductivity and Its Applications **412-14,** 976 (2004).

5. H. Safar *et al.*, Physical Review B **52,** R9875-R9878 (1995).

6. H. Douwes, P. H. Kes, C. Gerber, and J. Mannhart, Cryogenics **33,** 486 (1993).

7. L. Civale *et al.*, Applied Physics Letters **84,** 2121 (2004).

8. A. A. Abrikosov, A. I. Buzdin, M. L. Kulic, and D. A. Kuptsov, Superconductor Science & Technology **1,** 260 (1989).

9. W. Belzig, C. Bruder, and M. Sigrist, Physical Review Letters **80,** 4285 (1998).

10. G. Deutscher and K. A. Muller, Physical Review Letters **59,** 1745 (1987).

11. J. H. Durrell *et al.*, Physical Review Letters **90,** 246006 (2003).

12. A. Gurevich and L. D. Cooley, Physical Review B **50,** 13563 (1994).

13. E. M. Gyorgy *et al.*, Applied Physics Letters **56,** 2465 (1990).

14. W. K. Kwok *et al.*, Physical Review Letters **64,** 966 (1990).

15. J. Z. Liu, Y. X. Jia, R. N. Shelton, and M. J. Fluss, Physical Review Letters **66,** 1354 (1991).

16. M. Oussena *et al.*, Physical Review Letters **76,** 2559 (1996).

17. H. Safar *et al.*, Applied Physics Letters **68,** 1853 (1996).

18. M. Oussena *et al.*, Physical Review B **51,** 1389 (1995).

19. V. K. Vlaskovlasov *et al.*, Physical Review Letters **72,** 3246 (1994).

20. X. Obradors *et al.*, Superconductor Science & Technology **17,** 1055 (2004).

21. R. Herzog and J. E. Evetts, Review of Scientific Instruments **65,** 3574 (1994).

22. J. H. Durrell *et al.*, Physical Review B **70,** 214508 (2004).

23. J. M. Huijbregtse *et al.*, Superconductor Science & Technology **15,** 395 (2002).

24. M. G. Blamire and J. E. Evetts, Physical Review B **33,** 5131 (1986).

25. H. Suematsu, H. Okamura, S. Nagaya, and H. Yamauchi, Superconductor Science & Technology **12,** 274 (1999).

26. G. Blatter *et al.*, Reviews of Modern Physics **66,** 1125 (1994).